\documentclass[lettersize,journal]{IEEEtran}
\usepackage{amsmath,amsfonts}
\usepackage{algorithmic}
\usepackage{array}
\usepackage[caption=false,font=normalsize,labelfont=sf,textfont=sf]{subfig}
\usepackage{textcomp}
\usepackage{stfloats}
\usepackage{url}
\usepackage{verbatim}
\usepackage{graphicx}
\usepackage{color}
\def\BibTeX{{\rm B\kern-.05em{\sc i\kern-.025em b}\kern-.08em
    T\kern-.1667em\lower.7ex\hbox{E}\kern-.125emX}}
\usepackage{balance}
\newcommand{\phictr}{\tilde{\varphi}}

\begin{document}
\title{Deep-Space Optical Communication Receiver Based on Single Photon Coherent Beam Combination}
\author{
Antoni Mikos-Nuszkiewicz\IEEEauthorrefmark{1}\IEEEauthorrefmark{2},
Karol Łukanowski\IEEEauthorrefmark{1},
Konrad Banaszek\IEEEauthorrefmark{1}\IEEEauthorrefmark{2},
Marcin Jarzyna\IEEEauthorrefmark{1}
\\[1ex]
\IEEEauthorblockA{\IEEEauthorrefmark{1}Centre for Quantum Optical Technologies, Centre of New Technologies, University of Warsaw, Banacha 2c, 02-097 Warsaw, Poland}\\
\IEEEauthorblockA{\IEEEauthorrefmark{2}Faculty of Physics, University of Warsaw, Pasteura 5, 02-093 Warsaw, Poland}
}

\maketitle

\begin{abstract}
We introduce an alternative receiver architecture for deep-space optical communication, in which a single large aperture is replaced by an array of smaller ones with outputs combined coherently, employing phase stabilization based on photon counting events. We show that it allows to increase the signal to noise ratio, thus potentially attaining higher information transmission rates in the regime of large noise, typical for daytime communication. We analyze its practical performance by simulating pulse position modulation-based communication from the recently launched \emph{Psyche} mission. Under nighttime conditions the achieved performance is comparable to that offered by a single large aperture, whereas in daytime conditions the single photon coherent beam combination architecture provides an advantage in the information transmission rate.
\end{abstract}
\begin{IEEEkeywords}
coherent beam combination, pulse position modulation, deep space optical communication, photon-starved communication.
\end{IEEEkeywords}

\section{Introduction}
In recent years, free-space communication has been extended from traditional radio-frequency (RF) systems toward optical links \cite{trichiliRoadmapFreeSpace2020, hemmatiDeepSpaceOpticalCommunications2011, karmousHowCanOptical2025, ivanovReviewDeepSpace2025a}. This transition is driven primarily by the much shorter wavelengths and thus higher frequencies of optical carriers, which enable orders-of-magnitude increases in achievable bandwidth and information throughput \cite{williamsRFOpticalCommunications2020}. For short and medium-range applications, free-space optical (FSO) systems offer reduced size, weight, and power (SWaP) requirements \cite{williamsRFOpticalCommunications2020}, enhanced physical-layer security \cite{lopez-martinezPhysicalLayerSecurityFreeSpace2015}, and freedom from spectrum licensing constraints \cite{khalighiSurveyFreeSpace2014}. At the same time, the much lower beam divergence of optical signals opens the possibility of communication over distances that are challenging for traditional RF systems, making optical links a compelling candidate for future deep-space missions \cite{zwolinskiRangeDependenceOptical2018}. On the other hand, FSO introduces multiple layers of complexity, including stringent requirements on acquisition, tracking, and pointing stability, as well as limited link availability due to cloud blockage, all of which must be mitigated to enable reliable deployment \cite{kaushalOpticalCommunicationSpace2017}. To date, numerous FSO link demonstrations between ground stations and satellites in Earth orbit, as well as between satellites, have been reported, including SILEX \cite{oppenhauserEuropeanSILEXProject1990}, EDRS \cite{bohmerLaserCommunicationTerminals2012}, LCRD \cite{israelEarlyResultsNASAs2023}, and TBIRD \cite{schielerRecentOnOrbitResults2023}. Only a few optical links have so far been established at deep-space distances, where the photon-starved regime becomes the norm. An important early milestone was the Lunar Laser Communication Demonstration (LLCD), which provided bidirectional optical links between a lunar-orbiting spacecraft and Earth \cite{borosonOverviewResultsLunar2014}. More recently, the Deep Space Optical Communications (DSOC) experiment on board NASA’s \emph{Psyche} mission has achieved the farthest optical link to date between Earth and a deep-space spacecraft \cite{velascoDeepSpaceOptical2024}. In its latest demonstrations, \emph{Psyche} sustained downlink data rates on the order of $\sim 8$~Mbit/s from distances approaching $3$~AU, i.e., larger than the maximum Earth–Mars separation \cite{wrightDownlinkUplinkLaser2025}.

Even though DSOC offers substantially higher transmission rates than RF systems for the same transmitter power, inevitably at some range, the optical power collected by the receiver on the ground becomes extremely weak—ultimately reaching the single-photon regime. The most obvious ways to increase the average number of detected photons are to raise the transmitter optical power, enlarge its aperture, or, on the receiving side, to increase the diameter of the ground-based collecting telescope. Since reducing the SWaP budget of a space mission is essential, increasing the size of the ground-based receiving aperture remains the most practical option for extending optical communication links to the greatest achievable distances and highest rates. However, the cost of a telescope increases significantly faster than its collecting area \cite{belleScalingRelationshipTelescope2004}, especially when adaptive optics (AO) is required. An alternative approach would be to employ an array of smaller telescopes whose outputs are coherently combined, which so far has only been considered in scenarios where the received optical power vastly exceeds the photon-starved regime of DSOC~\cite{yangMultiapertureAllfiberActive2017, weyrauchExperimentalDemonstrationCoherent2011, larssonCoherentCombiningLowpower2022b, Sacchi:24}. Beyond the potential cost reduction, this approach might also be potentially more scalable, would offer simpler maintenance, and, as shown in this paper, in some instances it could improve the signal-to-noise ratio (SNR). A particularly promising feature is that coherent beam combination (CBC) yields a single coherent output beam, which enables advanced signal processing and may allow the use of future quantum-enhanced strategies for further performance gains, such as quantum pulse gating \cite{ecksteinQuantumPulseGate2011, Jarzyna2024}. 

In this work, we present a comprehensive study of single photon coherent combination of multiple beams in a multi-stage architecture which we designate as \emph{cascaded CBC}. The model includes wavefront distortion by atmospheric turbulence, where we account for both phase and intensity fluctuations, as well as the presence of background noise. We show how cascaded CBC can be used as a ground station receiver for communication with deep-space missions, based on the pulse position modulation (PPM) format for information encoding~\cite{xuPulsePositionModulation2009}. In the results section, we demonstrate how the performance of a cascaded CBC system is affected by two key parameters: phase diffusion and background noise level. We then analyze how different choices of the number of apertures and their individual sizes influence the overall system efficiency. Subsequently, we show that cascaded CBC scheme increases SNR by reducing the amount of background noise in the output beam at each combination stage. Finally, we present a realistic nighttime and daytime communication simulation between a deep-space transmitter and an Earth-based cascaded CBC receiver, modelling the conditions of the \emph{Psyche} mission and the Helmos Observatory. We observe that the CBC approach provides comparable information throughputs to a single large telescope case in nighttime conditions --- however, for daylight communication scenarios with significantly increased background noise, CBC demonstrates a clear advantage resulting from its noise filtering capability.

\section{Coherent beam combination}
In this section we discuss the techniques and assumptions underlying our scheme of coherent beam combination. In Sec.~\ref{sec:CBC} we cover the basic building block, coherent combination of two incoming photon-starved beams. Next, in Sec.~\ref{sec:limitations} we discuss the main hurdles faced by the scheme: phase fluctuation, intensity fluctuation, and background noise. Finally, in Sec.~\ref{sec:cascaded_cbc} we expand the binary model by considering the combination of multiple beams in a cascaded architecture. 

\begin{figure}[t]
    \centering
    \includegraphics[width=1\linewidth]{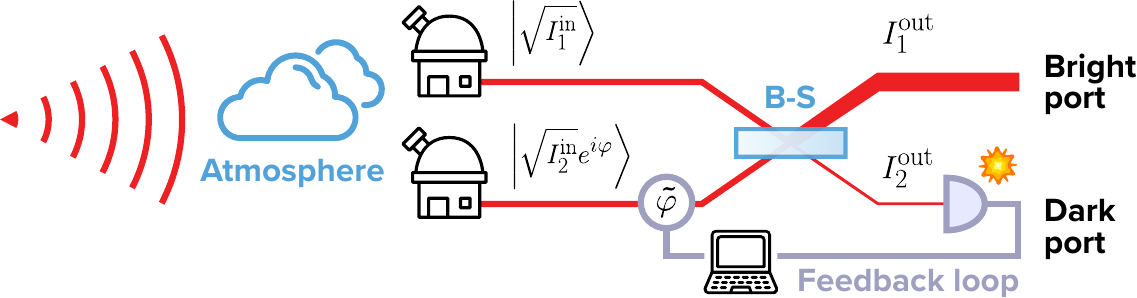}
    \caption{\textbf{Binary coherent beam combination in the photon-starved regime}. The optical signal is received by two separate apertures resulting in two light beams that are then interfered on a $50\mathpunct{:}\!50$ beam splitter, whose dark port is monitored by a photodetector. To ensure maximum amount of light leaving the scheme through the bright port, discrete detected photocounts are used to reduce the phase difference between the input beams --- which may result, for instance, from turbulent propagation in the atmosphere --- through a feedback loop. }
    \label{fig:scheme}
\end{figure}

\subsection{Binary coherent beam combination in the photon-starved regime} \label{sec:CBC}
The central concept underlying our results is a technique for achieving efficient coherent combination of two beams in photon-starved conditions which has been sketched in a simplified noise-free scenario in~\cite{mikos-nuszkiewiczBayesianApproachCoherent2024b}. Consider an incoming signal that has undergone diffraction, causing its spatial profile to expand, as well as atmospheric turbulence inducing wavefront distortion. The elementary CBC module visualized in~Fig.~\ref{fig:scheme} has two receiving apertures that collect spatially separate parts of this signal resulting in two beams that are then directed to the input ports of an interferometer. The goal of the scheme is to coherently combine the two beams, which is equivalent to achieving constructive interference in the bright output port and destructive interference in the dark one. In the simplest case, one may assume that the signal collected by each of the two apertures is described by quantum-optical coherent states with equal intensities $I^\text{in}$ and a phase difference $\varphi$. The two light beams interfere on a $50\mathpunct{:}\!50$ beam splitter. In the case of zero phase difference, $\varphi = 0$, all the light is directed into the bright port which results in perfect beam combination, output intensity $2 I^\text{in}$, and, at the same time, a lack of signal in the dark port. If, however, $\varphi \neq 0$, some part of light will leak out of the dark port yielding non-ideal CBC. To remedy that, a single photon detector (SPD) is placed in the dark port, whose goal is to indicate the presence of a  nonzero phase difference. Then, based on the registered photocount statistics, the necessary phase correction $\phictr$ is estimated and applied to one of the input beams through a feedback loop in order to reduce the phase delay for subsequent light pulses.

Crucially, due to the discrete nature of photons in the incoming signal, the photodetector in the dark port registers a count ($k=1$) or not ($k=0$) during time step $\Delta t$ only probabilistically, with respective probabilities
\begin{equation}\label{eq:prob_id}
p(k=1|\varphi)=1-e^{-I^{\text{out}}_2 \Delta t},\quad p(k=0|\varphi)=e^{-I^{\text{out}}_2 \Delta t},
\end{equation}
where
\begin{equation}\label{eq:iout}
I^\text{out}_{1,2} = I^\text{in} \left[1\pm \cos \left(\varphi - \phictr\right)\right]
\end{equation}
are the intensities of light respectively in the bright and dark port, as indicated in Fig.~\ref{fig:scheme}. The feedback algorithm that determines the optimal phase shift $\phictr$ from the detector’s click history is based on a maximum a posteriori (MAP) phase estimator, described in detail in \cite{mikos-nuszkiewiczBayesianApproachCoherent2024b}. Its goal is to eliminate clicks in the dark port which ideally indicates destructive interference with $\phictr = \varphi$, and hence constructive interference in the bright port. The efficiency $\eta$ of an elementary CBC module in Fig.~\ref{fig:scheme} can be defined simply as the ratio of the intensity leaving the system through the bright output port to the sum of the two input intensities, $\eta = I^\text{out}_{1} / (2 I^\text{in})$.

\subsection{Limiting factors for coherent beam combination} \label{sec:limitations}
In practice the efficiency of CBC is hindered by various realistic imperfections, such as the fluctuation of intensities and the phase difference between the beams, background noise or non-ideal detectors experiencing e.g. dark counts. Below we briefly discuss the most important effects that limit the CBC performance.
\subsubsection{Phase fluctuation}
\label{sec:phase_fluct}
The fundamental challenge in coherent beam combination of beams propagating through a turbulent medium, such as the atmosphere, is the randomly fluctuating relative phase difference between them. Specifically, for points separated by more than the so-called Fried parameter~\cite{friedOpticalResolutionRandomly1966a}, the phases are effectively mutually independent. In this work, we assume that our receiving apertures are separated by more than the Fried parameter and model phase fluctuation as a Gaussian random walk \cite{zhangExperimentalStudiesPhase2015}, where during each time step $\Delta t$ the phase difference changes by a small value $\Delta\varphi$ drawn from a normal distribution
\begin{equation}\label{eq:prob_phi}
    \Delta\varphi\sim\mathcal{N}(0,\sigma^2),
    \end{equation}
with zero mean and variance
\begin{equation}
\sigma^2 = 2 D_p \Delta t,
\end{equation}
where $D_p$ is the diffusion coefficient. Hence, the problem of efficient CBC becomes equivalent to optimally tracking the incoming phase $\varphi$ with a controlled phase $\phictr$ based on the history of photodetections in the dark port. 

\subsubsection{Intensity fluctuation}
\label{sec:intensity_fluct}
In the previous section, Sec.~\ref{sec:CBC}, we have assumed that the intensities of the incoming beams are equal and constant. If they differ and read at the two inputs $I^\text{in}_1$ and $I^\text{in}_2$ respectively, the output intensities are given by
\begin{align}
    I^{\text{out}}_{1,2}(\varphi) = \frac{1}{2}\left[I^{\text{in}}_{1}+I^{\text{in}}_{2}\pm2\sqrt{I^{\text{in}}_{1}I^{\text{in}}_{2}}\cos\left(\varphi-\phictr\right)\right] 
    \label{Eq:intensities}
\end{align}
which leads to a nonzero intensity level in the dark port even in the absence of a phase difference. In order to remedy this issue, one can replace the $50\mathpunct{:}50$ beam splitter in Fig.~\ref{fig:scheme} by one with reflectivity
\begin{align}
R = \frac{I^\text{in}_2}{I^\text{in}_1 + I^\text{in}_2}.    \label{eq:reflectivity}
\end{align}
This reduces the problem to the case of equal intensities, that is, if one is able to attain $\phictr = \varphi$, all of the light will leave the setup through the bright output port. The definition of combination efficiency naturally translates to $\eta = I^{\text{out}}_{1} / (I^\text{in}_1 + I^\text{in}_2)$.

The situation becomes more challenging when the incoming intensities additionally fluctuate in time, as their instantaneous values are unknown and the beam splitter reflectivity cannot be easily adjusted. A possible solution is to tap a small portion of both beams before the CBC stage and estimate their intensities online which would enable real-time adjustment of the main beam splitter's reflectivity. This can be achieved e.g. by using two auxiliary beam splitters with low reflectivity of around $\sim \! 1\%$, with photodetectors placed at their secondary ports. The drawback of this approach is a constant loss of the fraction $R$ of the output intensity. Moreover, when the beams are photon-starved, a significant integration time is required to accurately estimate their intensities. Therefore in this work we adopted the simple strategy with a constant, balanced $50\mathpunct{:}50$ beam splitter. 

Following the literature \cite{rytovPrinciplesStatisticalRadiophysics1988, andrewsLaserBeamPropagation2005}, we assume that due to Rytov theory, the intensities fluctuate weakly according to a lognormal random walk without a drift \cite{shreveStochasticCalculusFinance},
\begin{equation}\label{eq:prob_I}    
\log\!\left(I^\text{in}/I^{\text{in}}_0\right) \sim \mathcal{N}\!\left(0, \sigma_I^2\right),
\end{equation}
where $\log(I^\text{in}/I^{\text{in}}_0)$ denotes the logarithm of the input intensity relative to its mean value $I^{\text{in}}_0$ and the variance $\sigma_I^2$ is related to the intensity diffusion $D_I$ as $\sigma_I^2=2D_I\Delta t$. This model ensures that the input intensity fluctuates around its initial value $I_0^\text{in}$ while remaining strictly positive.
\subsubsection{Background noise}
\label{sec:background_noise}
A crucial element of the receiver is a spectral filter that cuts out the noise outside the frequency range occupied by the signal. However, even if the filtering used in the telescopes is highly efficient, background noise still impacts the incoming light in most scenarios \cite{majumdarFreeSpaceLaserCommunications2008, Raymer2020}. This is a particularly significant factor in daylight communication, as the Sun emits strongly across the entire optical spectrum. We assume that photons of the background radiation are incoherent and uncorrelated with the signal. Therefore, regardless of the controlled phase $\phictr$, the noise will always split equally at the beam splitter. Importantly, this implies that only half of the total noise collected by both telescopes leaves the CBC system through the bright port, while the remaining part leads to a constant light intensity in the dark port which can mislead the phase estimator with spurious clicks. Note that the latter effect is caused also by the detector noise, e.g., dark counts.

The background noise photon flux collected by a telescope can be estimated as~\cite{riekeDetectionLightUltraviolet2003}
\begin{equation}
I_\text{noise} = \frac{A \, L(\lambda)\, \Delta L\, \Omega }{h c / \lambda} =\frac{A \, \alpha_b(\lambda)}{h c / \lambda}, \label{eq:noise}
\end{equation}
where $A$ is the telescope aperture area, $\Delta L$ is the linewidth of the narrowband filter in place, $\Omega$ is the telescope’s field of view in steradians, and $L(\lambda)$ is the sky radiance, which for $\lambda = 1550\,\text{nm}$ typically ranges between $4 \times 10^{-5}$ and $10^{-4}\,\text{W}\,\text{m}^{-2}\,\text{sr}^{-1}\,\mu\text{m}^{-1}$ under nighttime conditions~\cite{ivanovReviewDeepSpace2025a}. The quantities in Eq.~\eqref{eq:noise} can be combined into the background noise power spatial density $\alpha_b(\lambda)=L(\lambda)\Delta L\Omega$ which characterizes the noise density received by a particular telescope.

\subsubsection{MAP estimator} To compute the MAP phase estimator in the case of phase fluctuations one follows the procedure described in \cite{mikos-nuszkiewiczBayesianApproachCoherent2024b}. Due to intensity fluctuations and background noise we consider here, one needs to modify the photon detection probability in the dark port, which is now expressed as
\begin{equation}
p_d(k_i=1|\varphi_i) = 1 - \exp\!\big[-(I^\text{out}_2(\varphi_i) + I_\text{noise})\,\Delta t \big],    \label{eq:prob_detection}
\end{equation}
where $k_i = 1$ denotes the detection of a photon in the $i$-th time step of duration $\Delta t$, and $I^\text{out}_2(\varphi_i)$ is the intensity at the dark port corresponding to the phase difference $\varphi_i$, as defined in Eq.~\eqref{Eq:intensities}. The MAP estimator of phase difference in the $i+1$-th step can then be obtained as
\begin{equation}\label{eq:estimator}
    \phictr_{i+1}=\arg\max_{\theta} p(\theta=\varphi_i|k_i,k_{i-1},\dots,k_1),
\end{equation}
where $p(\theta|k_i,k_{i-1},\dots k_1)$ denotes the probability of observing a phase difference $\theta$ and can be obtained from Eqs.~(\ref{eq:prob_phi}), (\ref{eq:prob_I}), and (\ref{eq:prob_detection}) by repeated application of Bayes' rule. Importantly, Eq.~\eqref{eq:estimator} does not specify the sign of the estimated phase difference, only its absolute value. However, when the sign is chosen erroneously, the corrected total phase difference becomes twice as large, resulting in a vastly increased probability of photodetection in the next step. Since the MAP estimator includes the history of detector clicks, it therefore quickly corrects the sign of the estimated phase if such immediate detection events occur.
\subsection{Cascaded coherent beam combination}
\label{sec:cascaded_cbc}
\begin{figure*}
    \centering
    \includegraphics[width=1\linewidth]{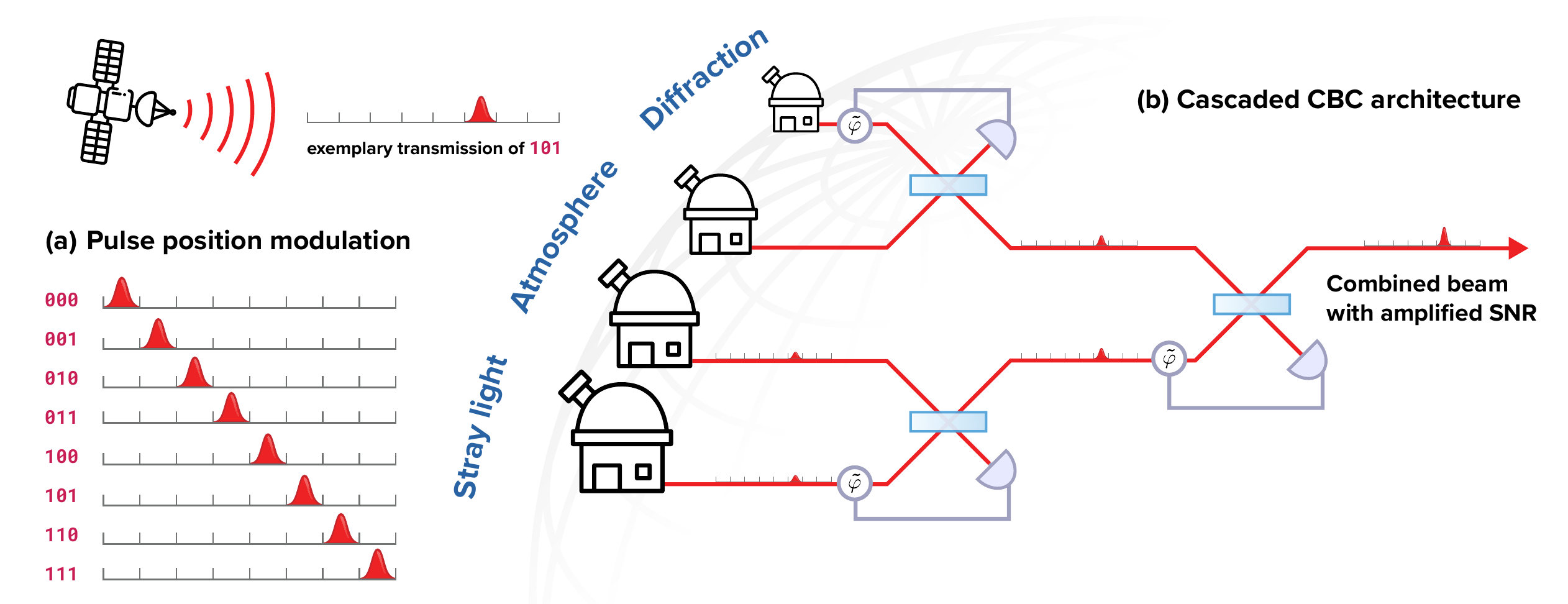}
    \caption{\textbf{Scheme of the proposed deep-space optical communication reception} based on a pulse position modulation format (a) received by an exemplary array of $4$ telescopes with coherently combined beams in the cascaded scheme (b). }
    \label{fig:ppm_CBC}
\end{figure*}

The elementary two-beam CBC schemes from Sec.~\ref{sec:CBC} can be further combined with each other in a tree-like cascade pictured in Fig.~\ref{fig:ppm_CBC}b). It is convenient to assume that there are in total $N = 2^n$ input beams, where $n$ denotes the number of vertical layers of the tree which we call \emph{stages}. In the first stage, $N$ beams are combined in pairs, resulting in $N/2$ output beams. These are then combined again with each other in pairs in the next stage, resulting in $N/4$ outputs, and so on, until only a single output beam remains. If the number of input beams $N$ is not a power of two, the beams can be combined by introducing unbalanced beam splitters with reflectivities given by Eq.~\eqref{eq:reflectivity} whenever the output of a beam pair is to be interfered with a previously unpaired beam. In this way, the number of combinations requiring unbalanced beam splitters is minimized.

The main difficulty of the multi-stage combination is that each CBC segment operates with a non-unit efficiency, and occasional phase estimation errors can cause temporary outages of the output beam, leading to sudden intensity drops in one of the input beams for the next stage. Such errors propagate through subsequent combinations, further degrading the cascaded CBC performance. Note that even seemingly high per-stage efficiency can significantly reduce the final output intensity when applied across many stages. For example, if the efficiency of a two-beam CBC is $95\%$ and there are $n = 7$ stages, the overall efficiency drops to at most $(95\%)^7 \approx 70\%$.

An important feature mentioned in Sec.~\ref{sec:background_noise} is that CBC of two beams in the presence of background noise results in an output beam in the bright port that posses only half on the total noise received by the two apertures. A profound consequence of this fact is that, in principle, in a cascaded CBC scheme the signal-to-noise ratio (SNR) can nearly double at each stage. Specifically, if at the $j$-th stage each beam has intensity approximately equal to $I_j$, background noise intensity $I_\text{noise}$, and combination efficiency $\eta$, one obtains
\begin{equation}
    \text{SNR}_{j+1} = \frac{2\eta I_j}{I_\text{noise}} = 2\eta\,\text{SNR}_j.
\end{equation}
Hence, for CBC of $N = 2^n$ beams, assuming the initial SNR of each received input beam is equal and given by $\text{SNR}_0$, the output beam of the multi-CBC scheme is characterized by SNR equal to
\begin{equation}
\text{SNR}_\text{out} = (2\eta)^n \text{SNR}_0.
\end{equation}
This effect can improve the efficiency of CBC at later stages, as an increased signal level compared to noise allows for better phase estimation. Moreover, the repeated SNR reduction --- even at the cost of attenuating the signal by $\eta$ at each stage --- may be useful for information decoding, especially in more noisy conditions typical for e.g. daytime communication.

\section{Deep-space optical communication}\label{sec:comm}
In this section we discuss the methodology behind the simulation of deep-space optical communication, in which we employ cascaded CBC of multiple beams in place of a single large receiving aperture. Specifically, Sec.~\ref{sec:ppm} describes the technique of pulse position modulation (PPM) with serially-concatenated forward error correction. The calculation of the received signal strength is described in Sec.~\ref{sec:link_equation}. Finally, Sec.~\ref{sec:comm_sim} covers the simulation of CBC for PPM reception.

\subsection{Serially Concatenated Pulse Position Modulation}
\label{sec:ppm}
Due to the exceedingly large distances traversed by the signal transmitted from deep-space missions to Earth and hence strong attenuation, DSOC modulation requires high photon information efficiency (PIE) defined as the ratio of transmitted bits to the number of photons received. The customary format of choice is pulse position modulation (PPM)~\cite{hemmatiDeepSpaceOpticalCommunications2011, Hemmati2006, Boroson2004} in which the transmission consists of a series of time frames depicted schematically in Fig.~\ref{fig:ppm_CBC}(a). Each PPM frame with modulation order $M$ is divided into $M$ slots and a pulse of light is prepared in exactly one slot, the others left empty. The position of the pulse within the $M$ slots is specified by $\log_2 M$ bits, and it is therefore convenient for $M$ to be a power of two. By keeping the average transmitter power constant and subdividing a given time frame into more slots, PIE is increased at the cost of shortening the pulse duration and hence, higher peak to average power ratio~\cite{Jarzyna2024, Banaszek2019}. 

The goal of PPM reception on Earth is to read out the position of the pulse-bearing slot in each frame. In the basic approach, after synchronization with the incoming frames, slot-by-slot direct detection is employed. Correct slot identification is hindered by the appearance of erasures (no photons being detected in the whole frame, typical for the photon-starved conditions of DSOC channels) and errors (detections in the empty slots, caused by background noise or imperfect extinction ratio of the transmitter). To combat those effects and allow for error-free communication, forward error correction (FEC) is applied at the transmitter, in which the messages to be transmitted are first encoded into longer messages with added redundancy. Due to this additional structure, if erasures and errors appear but do not exceed a threshold defined by the encoding, the receiver is able to decode the transmission perfectly and retrieve the original information content. 

In our simulation, we employ a custom implementation of the serially-concatenated PPM (SCPPM) coding protocol, currently recommended for DSOC by the CCSDS~\cite{CCSDS142}, whose full encoding and decoding algorithm is delineated in~\cite{Moision2005}. The original message content expressed in bits is first appended with a cyclic redundancy check, i.e., a bit sequence that is a highly sensitive function of the message content, which is verified at the decoder output to determine whether the decoding was successful. The bits are then fed through a convolutional encoder whose output is a convolution of the original message with judiciously chosen binary generator polynomials. For each input bit, the number of output bits corresponds to the number of polynomials, and thus the rate of the convolutional code is the inverse of the latter value. At this point, the rate can be increased by applying code puncturing, which entails removing certain bits from the encoding according to a chosen puncture pattern. Next, the encoded and punctured bits are passed through an interleaver that permutes them in a way reversible at the decoder. The goal of interleaving is to protect against burst errors, i.e., events which erase or modify several bits in a row --- if they happen during transmission, after deinterleaving their effect is smoothed out across the whole encoding, which makes it more probable for SCPPM to succeed in decoding. Finally, the interleaved bits pass through a rate-1 accumulator code (APPM) that translates them to PPM symbols. This completes the encoding and the encoded symbols are transmitted through the communication channel, with the possibility of the transmission being repeated a chosen number of times. The repeat factor is especially useful in severely photon-starved conditions, in which erasures dominate, and so multiple transmissions are needed for the detection of a photon in a given slot to happen at least once. Upon detection, the outputs are decoded in an iterative fashion using the Viterbi algorithm with two modules that exchange information: a decoder for the inner APPM code and for the outer convolutional code. The iterative decoding is performed until the cyclic redundancy check is satisfied, ensuring the message to be successfully decoded, or until a specified maximal number of iterations is reached and a decoding failure is announced. For specific details of our custom SCPPM code, see Results in Sec.~\ref{sec:cbc_ppm_sim}.

\subsection{Signal strength calculation with the link equation}
\label{sec:link_equation}
The average strength of the signal transmitted from deep space that reaches a detector on Earth can be calculated via the link equation~\cite{Moision2014},
\begin{equation}\label{eq:link}
    P_r = P_t \left( \frac{\pi d_t d_r}{4 R \lambda} \right)^2 \eta_{\text{tot}},
\end{equation}
where $P_r$ is the received signal power, $P_t$ the transmitted signal power, $d_r$ and $d_t$ the respective diameters of the receive and transmit apertures, $R$ is the distance between them, $\lambda$ the carrier wavelength, and $\eta_{\text{tot}}$ the overall system efficiency. The $1/R^2$ factor is responsible for the inverse-square-law attenuation of the signal power with distance, whereas other sources of loss are accounted for in the $\eta_{\text{tot}}$ parameter. These arise due to imperfect pointing and tracking, the effect of the atmosphere, detector quantum efficiency, additional loss in the transmitter and receiver, as well as the efficiency of CBC if it is implemented before detection.

\subsection{Communication simulation}
\label{sec:comm_sim}
We present now the approach for numerical simulation of CBC and subsequently demonstrate its application to communication using PPM, shown schematically in Fig.~\ref{fig:ppm_CBC}.

\subsubsection{Simulation of CBC}
We start by dividing the time axis into small steps of duration $\Delta t$. We choose a time unit $\Delta t = 1\,\text{ms}$, which corresponds to the timescale over which the phase of a beam propagating through the atmosphere does not change significantly. We assume $N = 2^n$ input beams. Each beam is characterized by an intensity and a phase that fluctuate independently according to lognormal and normal random walks, introduced in Eqs.~\eqref{eq:prob_I} and \eqref{eq:prob_phi} respectively, at each time step $\Delta t$, as described in Sec.~\ref{sec:limitations}. All intensities start from a fixed value $I_0^\text{in}$, and all initial phases are drawn randomly from the range $[-\pi, \pi]$. Additionally, every beam  carries background noise intensity $I_\text{noise}$ which is constant in time and calculated according to Eq.~\eqref{eq:noise}. The beams are combined pairwise and independently. For every pair at time step $i$, we draw whether a photodetection occurs in the dark port according to the probability given in Eq.~\eqref{eq:prob_detection}. We then estimate the phase difference using the Bayesian estimator described in Sec.~\ref{sec:CBC}, apply the correction $\phictr_i$ for the next time step, and compute the resulting intensity and phase. These values serve as inputs for the subsequent combination stage. This procedure is repeated $n$ times until the final combined beam is obtained. In the subsequent iteration in time, the process is repeated using the previously calculated controlled phases $\phictr$ for each combination pair.

\subsubsection{Simulation of PPM transmission}
To simulate communication, we encode a random message using the SCPPM protocol and generate a signal composed of short pulses with intensity $I_\text{pulse}$ in specific time bins defined by the PPM format. In the time slots without a pulse, the transmitted intensity equals $I_\text{pulse}$ multiplied by an extinction ratio $\textrm{ER}$. The signal is then attenuated to a level determined from the link-budget equation~\eqref{sec:link_equation} and collected by $N$ apertures. Next, we apply the CBC simulation described above and obtain a time-varying intensity of the combined beam. Since typically the CBC timescale can be up to $10^3$ times longer than the slot duration for PPM, the input intensity for CBC can be averaged over the interval $\Delta t$. Finally, a single-photon detector placed at the output of the cascaded CBC scheme is simulated. A bit value of $1$ is assigned to time slots in which a photon is detected and $0$ otherwise. The message is then decoded based on the detected bit sequence.

\subsubsection{Stage optimization in CBC} \label{sec:early}
\begin{figure}
    \centering
    \includegraphics[width=1\linewidth]{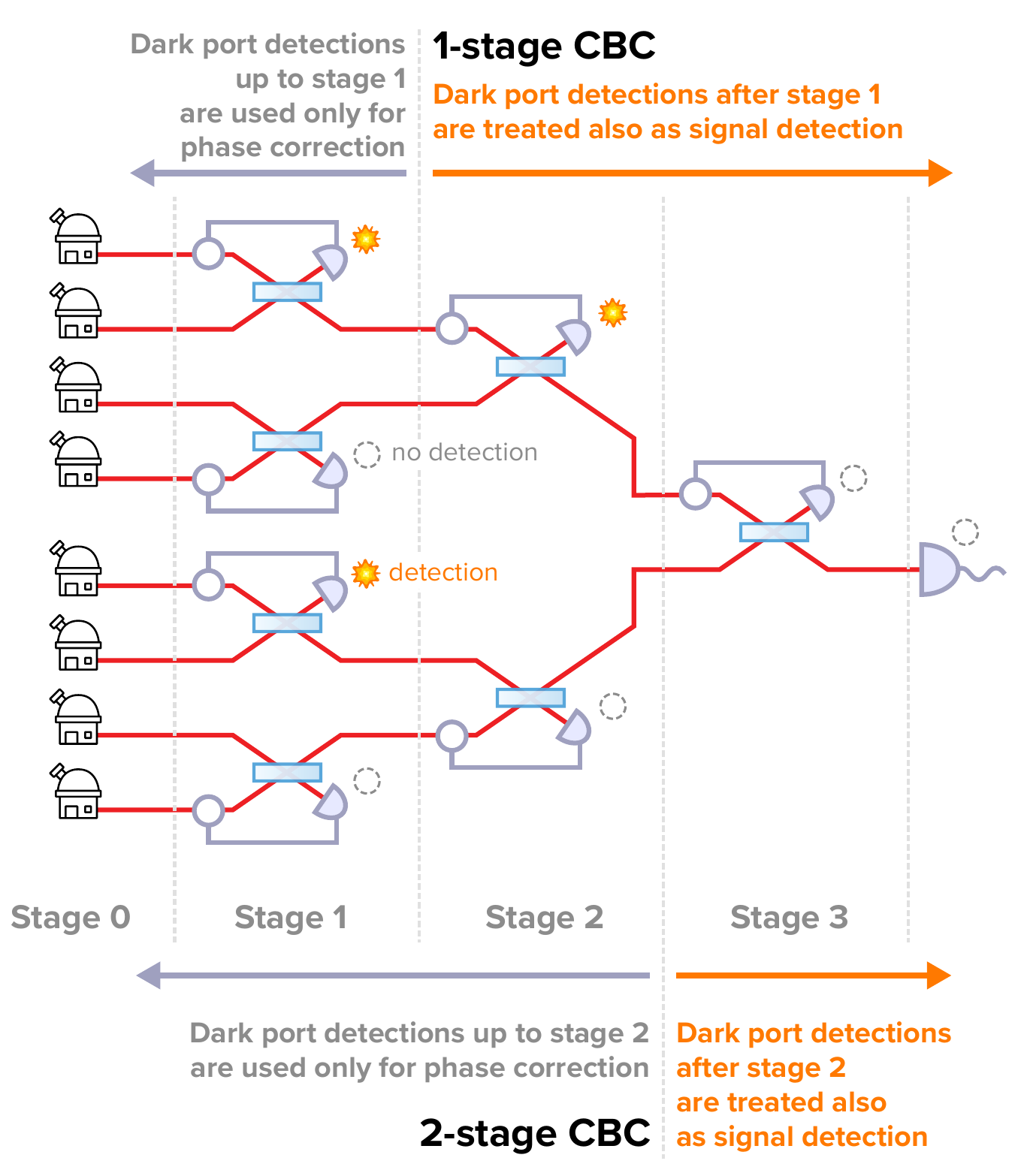}
    \caption{\textbf{Schematic representation of k-stage CBC} defined in Sec.~\ref{sec:early}, where despite performing full cascaded CBC, information decoding is optimized by the choice of a threshold stage $k$. Dark port clicks that happen after stage $k$ of cascaded CBC are treated as a detection of the signal. In the depicted example, 8 beams are combined and (no)-detection events that happened in the given communication slot are annotated next to their respective detectors. If 1-stage CBC is assumed, the slot is counted as occupied by the signal because of the detection in a dark port in stage 2 and despite a lack of detection in the final detector. On the other hand, if 2-stage CBC is assumed, the slot is considered empty as no detections happened from stage 3 onwards. The choice of the optimal $k$ for information decoding can be made in post-processing if one saves the information on individual slot-by-slot detector clicks.}
    \label{fig:k_CBC}
\end{figure}
Although full CBC of multiple beams into one output beam enables advanced processing of the output beam and a multifold increase of the SNR, in some cases it may actually be more beneficial for communication detected directly slot-by-slot, such as PPM, to measure the signal independently in each aperture without beam combination, or to perform direct detection after some number of combination stages but before the single output beam is produced. This stems from the fact that although each combination stage halves the total background noise remaining in the beams, it also attenuates the total signal intensity due to imperfect combination efficiency. Then, especially in low noise conditions, further removal of noise may be unnecessary, while the attenuation of signal can be detrimental to information transmission. This motivates us to introduce the notion of $k$-stage CBC, where the first $k$ stages of cascaded CBC are performed, after which the remaining beams are detected directly. Viewed this way, $n$-stage CBC corresponds to a full cascaded CBC of $2^n$ input beams into one output beam, and $0$-stage CBC entails direct detection of each incoming beam without any actual combination taking place.

However, it is important to note that the full cascaded CBC scheme naturally accomodates $k$-stage CBC for all $ 0 \leq k \leq n$, without the need for additional equipment such as photodetectors placed after the $k$-th stage. One can perform the full cascaded CBC process arriving at a single combined beam at the output, but still include the slot-by-slot photocount information from the dark ports to optimize information decoding. In this work, we adopt a simple recipe on how to operate $k$-stage CBC, pictured schematically in Fig.~\ref{fig:k_CBC}, in which we treat the dark output port detections in stages $> k$ as detections of the signal, while the photocounts from dark ports in stages up to $k$ are used only for phase difference correction. In other words, to answer whether a given time slot yields a detection, we check not only the final detector, but also whether \emph{any} photon was detected in \emph{any} dark port after the chosen threshold CBC stage $k$, treating all such detections equivalently. For each simulated communication scenario we may then choose an optimal $k$ in post-processing to be the one which allows for best performance of information decoding.

In fact, this approach could be further improved by specifically estimating the signal and noise intensities after each stage and finding how the photons detected at later stages are more likely to originate from the signal rather than from noise, compared to photons detected in the initial stages. Intuitively, the decoder could then give more credibility to those late-stage detections. We leave this research direction for future work.

\section{Results}
In this section we present the results of simulation of PPM communication with the cascaded CBC receiver. We begin with a discussion of the efficiency of cascaded CBC in terms of key parameters: phase diffusion $D_p$ and background noise strength $I_\text{noise}$. Then we discuss the potential gain from dividing a large aperture into smaller ones with coherently combined outputs, where we explain how to determine the optimal diameter of the small apertures and compare it to the case of a single large one. Subsequently, we show how cascaded CBC increases the SNR at each stage, providing an advantage over the single-aperture scenario under strong noise conditions. Finally, we present the attainable transmission rates for simulated downlink communication scenarios from the \emph{Psyche} spacecraft as a function of distance under nighttime and daytime conditions.

\begin{figure}[t]
    \centering
    \includegraphics[width=\linewidth]{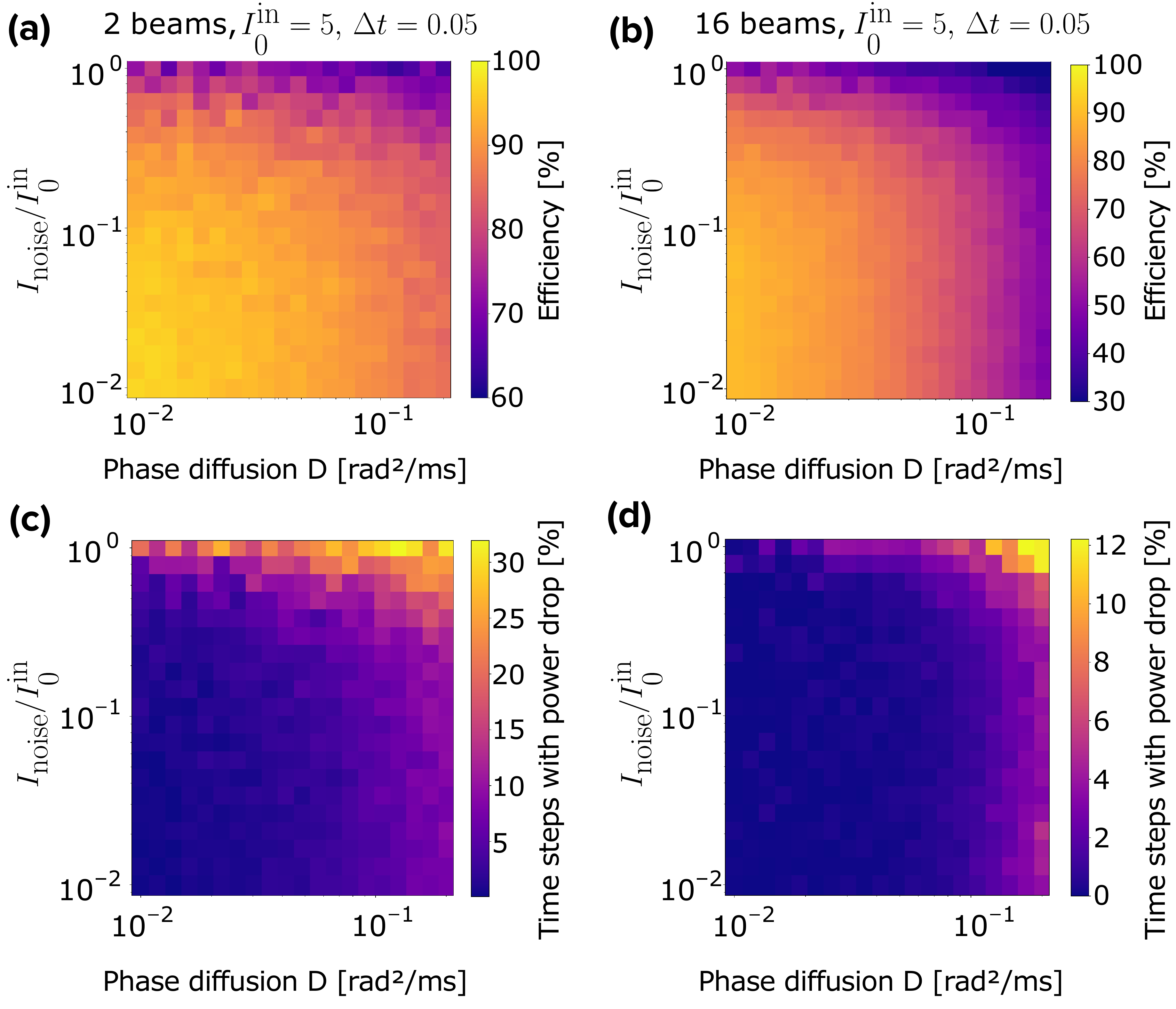}
    \caption{\textbf{Single photon coherent beam combination efficiency} for (a) $2$ and (b) $16$ input beams for $I_0^\text{in}=5$ [photons/ms],  $D_I=0, \Delta t=0.05$~ms as a function of phase diffusion parameter $D_p$ and the ratio of the the background noise and the total input intensity $I_\text{noise}/I_0^\text{in}$. (c), (d) Fraction of time steps for which final output intensity drops below $5\%$ of total initial value for respective cases.}
    \label{fig:CCBC}
\end{figure}

\subsection{Efficiency of cascaded CBC}\label{sec:efficiency}
To show the influence of phase diffusion and background noise on the efficiency of cascaded CBC, we simulate the combination of $N = 2$ and $N = 16$ beams for phase diffusion in the range $D_p \in [10^{-2}, 0.2]$ $\text{rad}^2\text{/ms}$ and for the ratio of background noise to initial intensity $I_\text{noise}/I_0^\text{in} \in [10^{-2}, 1]$. We assume a time step of $\Delta t = 0.05$ ms, an initial beam intensity $I_0^\text{in} = 5$ photons/ms, and no intensity fluctuation for simplicity, $D_I=0$. At each time step we calculate an instantaneous efficiency defined as the ratio of the current output intensity to the total input intensity. We define efficiency as an average of instantaneous efficiencies over a long time period, which we set to $10^4$ time steps. This is, according to the ergodic theorem \cite{birkhoffProofErgodicTheorem1931}, equivalent to taking the average over all possible realizations of the phase random walk, provided sufficiently long averaging times are taken. 

The simulated efficiency for $N=2$ and $N=16$ beams is plotted in Fig.~\ref{fig:CCBC}(a) and (b) respectively. It is seen that CBC efficiency decreases with increasing noise and phase diffusion. Interestingly, background noise starts to significantly affect the combination only when its strength approaches the signal. This leads to a rule of thumb: cascaded CBC is practical only when $I_\text{noise} < I_0^\text{in}$, where it consistently demonstrates high robustness against background noise. Next, it is seen that the efficiency of combining two beams must exceed $90\%$ to maintain high overall efficiency in the cascaded scheme, since it decreases exponentially with the number of combination stages. Importantly, although the efficiency of two-beam CBC in Fig.~\ref{fig:CCBC}(a) can be satisfactory, the output intensity may become too unstable for further combination stages. This can be quantified by the percentage of time steps in which an intensity outage occurs, defined as an instance in which the output intensity drops below $5\%$ of the total initial power, which can be seen in Fig.~\ref{fig:CCBC}(c), (d). When the number of outages becomes significant, it amplifies disturbances in subsequent CBC stages. As can be seen in Fig.~\ref{fig:CCBC}(c), for phase diffusion greater than $0.1$ $\text{rad}^2\text{/ms}$, outages appear in about $10\%$ of time steps during the combination of two beams, leading to a dramatic efficiency drop after four CBC stages for $N=16$ beams.

\subsection{Optimal number and diameter of individual apertures in an array} \label{sec:diameters}
The fundamental questions in designing a practical CBC receiver system are how many apertures to use and what their sizes should be. To address these, we assume a fixed phase diffusion parameter $D_p$ and an incident signal flux of $\Phi_0^\text{in}$ photons per millisecond per square meter. The key quantity to analyze is the average intensity $I^*$ needed to obtain the desired transmission rate which would be collected by a single aperture with diameter $d^*$, i.e. $I^*=\pi d^{*2}\Phi_0^\text{in}/4$. The lower limit on diameter $d_\text{CBC}$ of one of the small apertures from the CBC array is determined by the requirement that the intensity received by a single aperture $I_0^\text{in} = \pi d_\text{CBC}^2\Phi_0^\text{in}/4$ must be high enough for the CBC efficiency between two beams to reach at least $90\%$, as stated in Sec.~\ref{sec:efficiency}. Otherwise, the overall cascaded CBC efficiency $\eta_{\textrm{tot}}$ will decrease rapidly, preventing effective receiver scaling.  The second characteristic diameter value is equal to the Fried parameter \cite{friedOpticalResolutionRandomly1966a} which can reach over $20$~cm for good atmospheric conditions. The Fried parameter characterizes the spatial scale over which the wavefront remains phase-coherent. It is advantageous to keep the apertures diameters below this value to avoid the need for AO~\cite{tysonFieldGuideAdaptive2012} or to allow the use of simpler and less expensive methods such as tip-tilt correction~\cite{raoAstronomicalAdaptiveOptics2024}. 

\begin{figure}[t]
    \centering
    \includegraphics[width=1\linewidth]{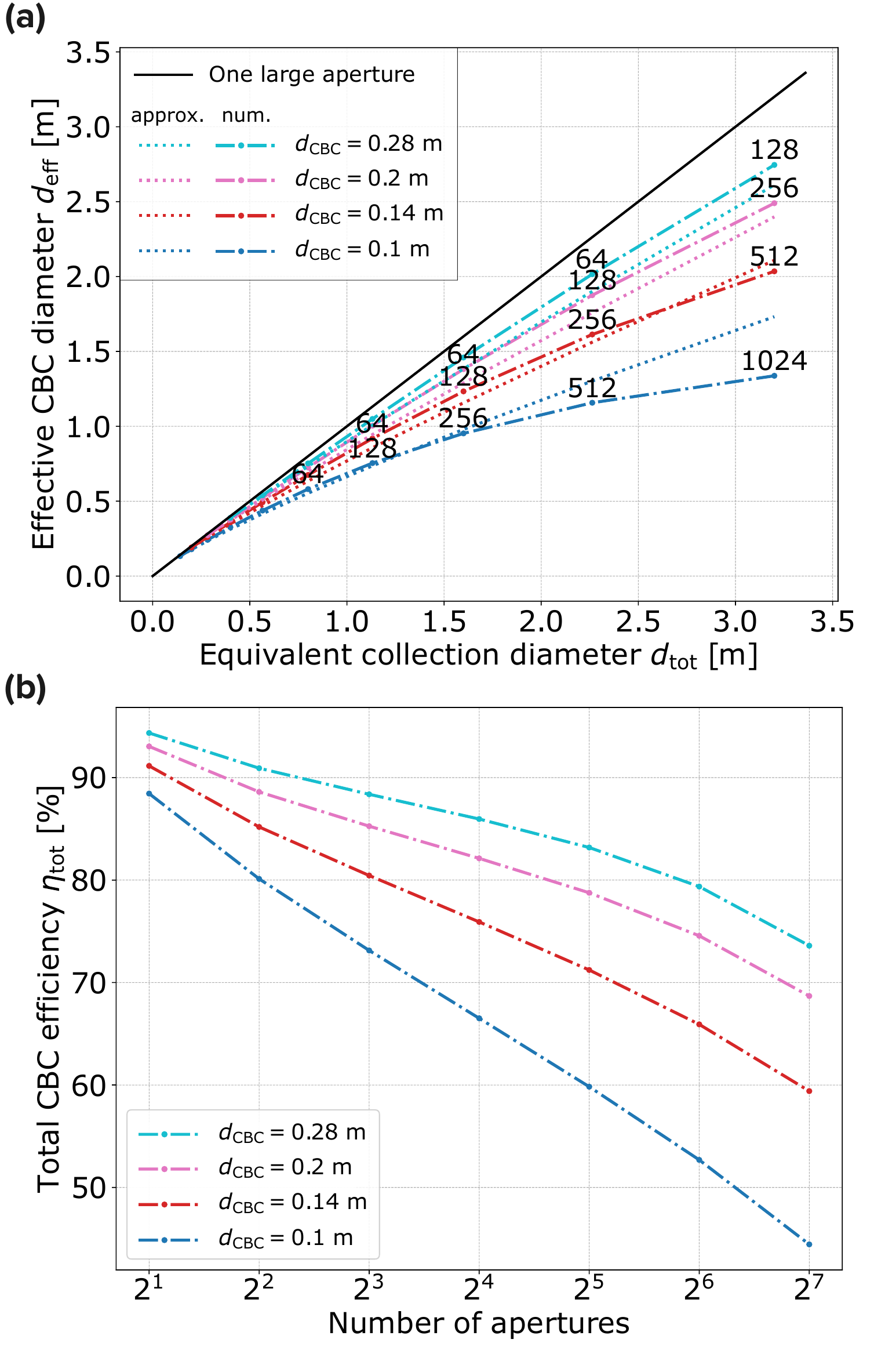}
    \caption{ (a) Diameter $d_{\textrm{eff}}$ of an effective single large aperture that collects the same optical power as the output of a CBC scheme with a given total collection area represented by its equivalent diameter $d_{\textrm{tot}}$ for different individual CBC apertures sizes $d_{\textrm{CBC}}$. Number of apertures $N$ is indicated by the number of CBC stages $n=\log_2 N$ above respective points and lines are drawn only to guide the eye. (b) Efficiency $\eta_{\textrm{tot}}$ of cascaded coherent beam combination as a function of the number of apertures for different diameters.}
    \label{fig:diameters}
\end{figure}

To provide guidance for selecting an appropriate diameter of CBC apertures one can calculate the size $d_{\textrm{eff}}$ of a single large aperture that is needed to attain the same level of output light intensity as a number $N$ of smaller apertures $d_{\textrm{CBC}}$ of a given size used for CBC
\begin{equation}\label{eq:diameter}
    d_{\textrm{eff}}=d_{\textrm{CBC}}\sqrt{\eta_{\textrm{tot}}N},
\end{equation}
where $\eta_{\textrm{tot}}$ denotes the total efficiency of a cascaded CBC system. Note that one needs to ensure also that $d^* \leq d_{\textrm{eff}}$, to successfully decode the message. Such comparison for various diameters is seen in Fig.~\ref{fig:diameters}~(a) for $D_p = 0.05$ $\text{rad}^2\text{/ms}$, $\Phi^\text{in}_0 = 344$~photons/$(\textrm{ms}\cdot \textrm{m}^2)$  which corresponds to the signal flux from the \emph{Psyche} mission at a distance of $1$~AU, under the assumptions listed in the Table~\ref{tab:communication}, background noise $I_\text{noise} = 0.1\cdot I_0^\text{in}$, and no intensity fluctuations. Specifically, the horizontal axis represents the diameter of an aperture whose collecting area equals the total area of $N$ smaller apertures $d_{\textrm{tot}}=d_{\textrm{CBC}}\sqrt{N}$, while the vertical axis shows the effective diameter $d_\textrm{eff}$ corresponding to a single large aperture that would collect the same total intensity as $N$ small apertures after cascaded CBC in \eqref{eq:diameter}. If the CBC was perfect, the curves would follow the diagonal black line, indicating that the same total collecting area yields the same intensity as a single large aperture of such area. It can be seen that as the diameter of the individual CBC apertures decreases, the curves deviate further from the diagonal. For example, the curve for $d_{\textrm{CBC}}=10$~cm apertures nearly saturates, meaning that for $d_\textrm{tot}\geq 2$~m doubling the number of apertures does not significantly increase the collected signal power due to the rapidly decreasing efficiency $\eta_{\textrm{tot}}$ in later combination stages, as is seen in Fig.~\ref{fig:diameters}~(b). This suggest that in these conditions aperture of size $10$~cm is sufficiently efficient for $d^* \sim 1$~m but becomes almost impractical for $d^* > 1.5$~m. When comparing CBC apertures of $20$~cm and $14$~cm diameter, the former offers approximately twice the collecting area. However, to reach comparable output power as e.g. the $d^*=2.28$~m telescope at the Helmos Observatory, one needs almost $2^7 = 128$ apertures of $20$~cm or nearly four times as many $14$~cm ones meaning the former strategy is the suitable choice.

In general, the relation between effective and total equivalent aperture diameters $d_{\textrm{eff}}$ and $d_{\textrm{tot}}$ is complicated. In order to give a guiding tool for the choice of CBC aperture size one can instead use an approximation of overall combination efficiency $\eta_\textrm{tot}$ assuming the efficiency of each CBC stage are the same and equal to the efficiency of combining two beams $\eta$, giving $\eta_\textrm{tot}\approx \eta^{\log_2 N}$. It is seen in Fig.~\ref{fig:diameters}(a) that such an approximation gives slightly lower values for a smaller number of combined beams. However, when the number of beams grows, and with it the number of stages in the cascade, the approximation overestimate the efficiency because it neglects the detrimental impact of outages in one stage on the input beams in subsequent CBC stages.

It is seen in Fig.~\ref{fig:diameters}(b) that efficiency decreases with the number of CBC stages and is highly sensitive to the efficiency of the first combination stage. A slight reduction in the initial efficiency leads to a drastic drop at later stages. Note that using smaller apertures lowers the efficiency of the first combination stage, while additional stages are required to achieve the same output signal intensity as with larger apertures. Consequently, arrays with too small individual CBC aperture size $d_{\textrm{CBC}}$ may become inefficient, depending on the required total signal collecting area.

The above approach may also be applied in scenarios where individual apertures, due to lower overall cascaded CBC efficiency, cannot be made smaller than the Fried parameter, particularly when detecting extremely low-power signals. For example, instead of constructing a single aperture with $d^*=10$~m, one could employ four $5$~m apertures. Although each of these apertures would require AO, the overall solution may still be economically advantageous.

Notably, because CBC is performed in the single-photon regime, it allows the use of smaller apertures than approaches that require stronger signals.
\subsection{The impact of noise on attainable information rate} \label{sec:res_SNR}
\begin{table}[t]
\centering
\caption{Simulation parameters used in Sec.~\ref{sec:res_SNR} to gauge the impact of noise on attainable information rates.}
\begin{tabular}{l c l}
\hline
\textbf{Symbol} & \textbf{Value} & \textbf{Description} \\
\hline
$T$ & $10^4$ & Number of CBC time steps \\
$M$ & $128$ & PPM rank \\
$\text{ER}$ & $10^{-3}$ & Extinction ratio \\
$\Delta t_\text{PPM}$ & $8\times10^{-6}$ & Duration of one PPM time slot [ms] \\
$N$ & $64$ & Number of combined beams \\
$D_p$ & $0.05$ & Phase diffusion coefficient [$\text{rad}^2\text{/ms}$] \\
$D_I$ & $0$ & Intensity fluctuation coefficient [$\text{ms}^{-1}$]\\
$\Delta t$ & $0.05$ & CBC simulation time step [ms] \\
$I_0^\text{in}$ & $30$ & Initial intensity [photons/ms] \\
\hline
\end{tabular}
\label{tab:SNR}
\end{table}
As explained in Sec.~\ref{sec:cascaded_cbc}, cascaded CBC eliminates half of the background noise from the output beam at each stage. However, at the same time, finite combination efficiency decreases the total number of detected signal photons. To assess how this trade-off affects available information rates, we simulated PPM-based communication with parameters listed in Table~\ref{tab:SNR} for a range of background noise values performing CBC with $64$ beams at the receiver side. The attainable communication rate is upper bounded by mutual information, which we have estimated from the simulation, see Appendix~\ref{app:MI}. Importantly, as noted in Sec.~\ref{sec:early}, one may choose the threshold stage of the cascaded CBC scheme after which all photocounts from the dark ports are used for decoding. Specifically, for each PPM time slot, if a photodetection event occurred in any of them, we assume it was occupied by a light pulse; otherwise we assume the slot was empty.

\begin{figure}[t]
    \centering
    \includegraphics[width=1\linewidth]{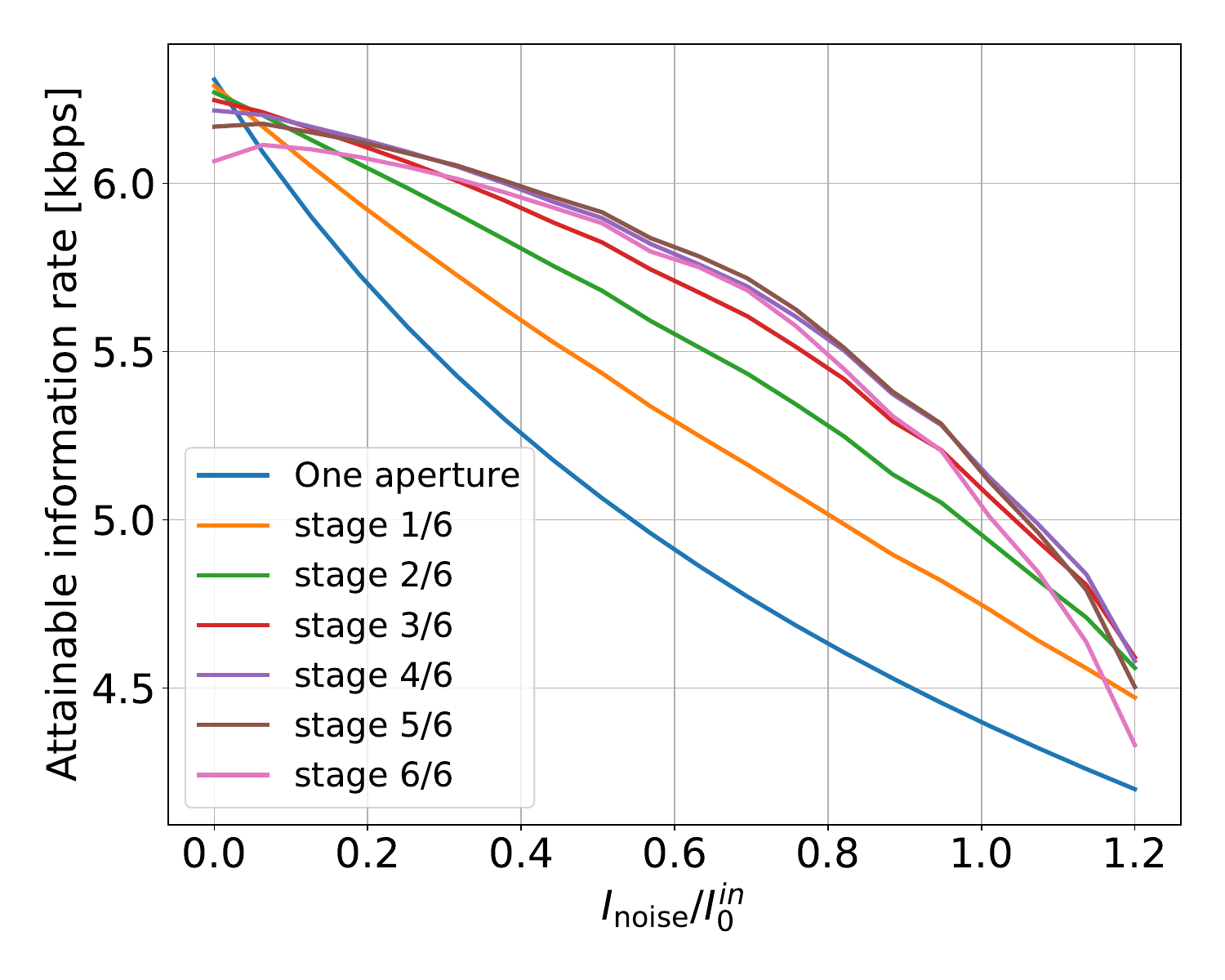}
    \caption{\textbf{Maximum communication rate for PPM and CBC of $64$ beams} with parameters listed in Table \ref{tab:SNR} as a function of the ratio between the background noise $I_\text{noise}$ and the signal intensity $I_0^\textrm{in}$. The curves correspond to different choices of the threshold combination stage after which all photons detected in the dark ports of the subsequent stages are included in the message decoding. The blue line corresponds to the scenario with one large aperture with area equal to the total area of all $64$ beams.}
    \label{fig:SNR}
\end{figure}

It is seen in Fig.~\ref{fig:SNR} that in a scenario with negligible background noise, $I_{\text{noise}} \to 0$, the most efficient approach is to account for photons from all the dark ports already starting from the first CBC stage. In the absence of dark counts, this is equivalent to a single large aperture with the same total collection area as the CBC array which means that in the regime of noiseless signal transmission CBC is not beneficial. This is because all the signal photons are in principle detected in the single-large-aperture scenario whereas CBC is intrinsically limited by its finite efficiency. However, adding even a small amount of noise to the system changes the behavior as spending some part of the signal from the first few stages on combining the beams becomes beneficial. This result clearly demonstrates that CBC approach is flexible across different background noise levels. For weak noise, the system can be treated as a set of independent small apertures with performance equivalent to a single large aperture with the same total area. For larger SNR, however, the scheme provides a clear advantage by offering significantly higher robustness to noise. Notably, as discussed in Sec.~\ref{sec:early}, the cascaded CBC architecture supports all of these operating regimes without requiring any physical modifications. Switching between them is achieved simply by optimizing over the combination stage used in post-processing.

\subsection{Downlink transmission simulation in nighttime and daytime conditions}
\label{sec:cbc_ppm_sim}
To estimate the performance of cascaded CBC in practical DSOC, we simulated downlink PPM transmission with SCPPM forward error correction implemented for information encoding and decoding, as described in Sec.~\ref{sec:comm}. We strove to model conditions similar to those experienced in the ongoing \emph{Psyche} mission, whose one of the goals is to demonstrate DSOC capabilities \cite{rielanderESAGroundInfrastructure2023b}. Unfortunately, a direct comparison is quite ambiguous. For one, two ground stations are envisioned for the mission, the Aristarchos telescope of $2.28\,$m in diameter, situated at the Helmos Observatory, and the $5.1\,$m Hale telescope at Palomar Observatory. Noise conditions may differ at the two locations and we are not aware of their reliable measurements. For Aristarchos, theoretical predictions of information throughput, assuming PPM of order 128 and 4 W of \emph{Psyche} spacecraft's transmitter power, are included in ref.~\cite{rielanderESAGroundInfrastructure2023b}. For Hale, ref.~\cite{wrightDownlinkUplinkLaser2025} reports a successful demonstration of downlink information transfer, but with PPM orders 16 and 32, 2W of transmitter power, and varying slot duration. Moreover, the degree of loss unrelated to distance, i.e. the $\eta_{\text{tot}}$ parameter in the link equation~\eqref{eq:link}, also varies between references: ref.~\cite{rielanderESAGroundInfrastructure2023b} assumes around 21 dB; the photon flux calculated from the bitrates and PIE reported in ref.~\cite{wrightDownlinkUplinkLaser2025} is consistent with a range of 7-14 dB; and the recent DSOC review of ref.~\cite{ivanovReviewDeepSpace2025a} cites a value of 14 dB. In the following, we have tried to retain as much generality as possible by indicating the signal and noise photon fluxes used in the simulations along with a consistent transmission distance calculated via the link equation~\eqref{eq:link}.

\begin{table}[t]
\centering
\caption{Simulation parameters used in Sec.~\ref{sec:cbc_ppm_sim} for real-condition nighttime and daylight Psyche-Helmos communication.}
\begin{tabular}{l c l}
\hline
\textbf{Symbol} & \textbf{Value} & \textbf{Description} \\
\hline
$M$ & $128$ & PPM rank \\
$\text{ER}$ & $10^{-3}$ & Extinction ratio \\
$\lambda$ & $1550$~nm & Transmit wavelength \\
$P_t$ & $4$~W & Average transmitted power \\
$d_t$ & $0.22$~m & Transmitter aperture diameter \\
$\Delta t_\text{PPM}$ & $8\times10^{-6}$ & Duration of one PPM time slot [ms] \\
$\eta_{\text{tot}}$ & $18.07$~dB & Distance-unrelated loss \\
$\alpha_b^d(\lambda)$ & $3.1\cdot 10^{-2}$ pW/$\text{m}^2$ & Daytime background noise \\
$\alpha_b^n(\lambda)$ & $3.1\cdot 10^{-5}$ pW/$\text{m}^2$ & Nighttime background noise \\
$N$ & $128$ & Number of combined beams \\
$d_\textrm{CBC}$ & $20$~cm & Diameter of CBC apertures \\
$D_p$ & $0.05$ & Phase diffusion coefficient [$\text{rad}^2\text{/ms}$] \\
$D_I$ & $0.05$ & Intensity fluctuation coefficient [$\text{ms}^{-1}$]\\
$\Delta t$ & $0.05$ & CBC simulation time step [ms] \\
\hline
\end{tabular}
\label{tab:communication}
\end{table}

The base SCPPM forward error correction algorithm was implemented according to the prescription of ref.~\cite{Moision2005}. A single message length was fixed to 15120 bits after convolutional encoding and interleaving, irrespective of the code rate. The PPM order was set to 128, so that each PPM symbol encoded $\log_2 128 = 7$ bits and a transmission of a single message consisted of $15120 / 7 = 2160$ symbols. A transmission simulation was deemed successful if 25 subsequent messages were received and decoded without error. The original SCPPM implementation was augmented to allow for finer control of the code rate. First, the convolutional encoder was given the option to choose from different sets of generator polynomials: the rate-1/2 $(5_8, 7_8)$, rate-1/3 $(5_8, 7_8, 5_8)$, as well as repeats of those set. Second, code puncturing was implemented, both with puncture patterns found in literature~\cite{Yasuda1984,Haccoun1989} and with custom ones determined heuristically. Finally, we allowed for transmission repeats, in which a single encoded message was transmitted more than once. After reception, the repeated slot sequences were translated to a single sequence according to the following rule: a slot was considered empty if it had no detections in each of its repeats; otherwise, if a photon was detected at least once in the repeats of the slot, it was marked as occupied. While this was akin to a crude repetition code, it turned out crucial in the severely photon-starved conditions at the highest transmission distances, where erasures dominated and repeats of transmissions were necessary to measure any photons at all in the pulse-bearing slots. For each transmission simulation, the convolutional code polynomials, puncture patterns, and number of transmission repeats were optimized over to achieve error-free communication at the highest possible rate.

\begin{figure}[t]
    \centering
    \includegraphics[width=1\linewidth]{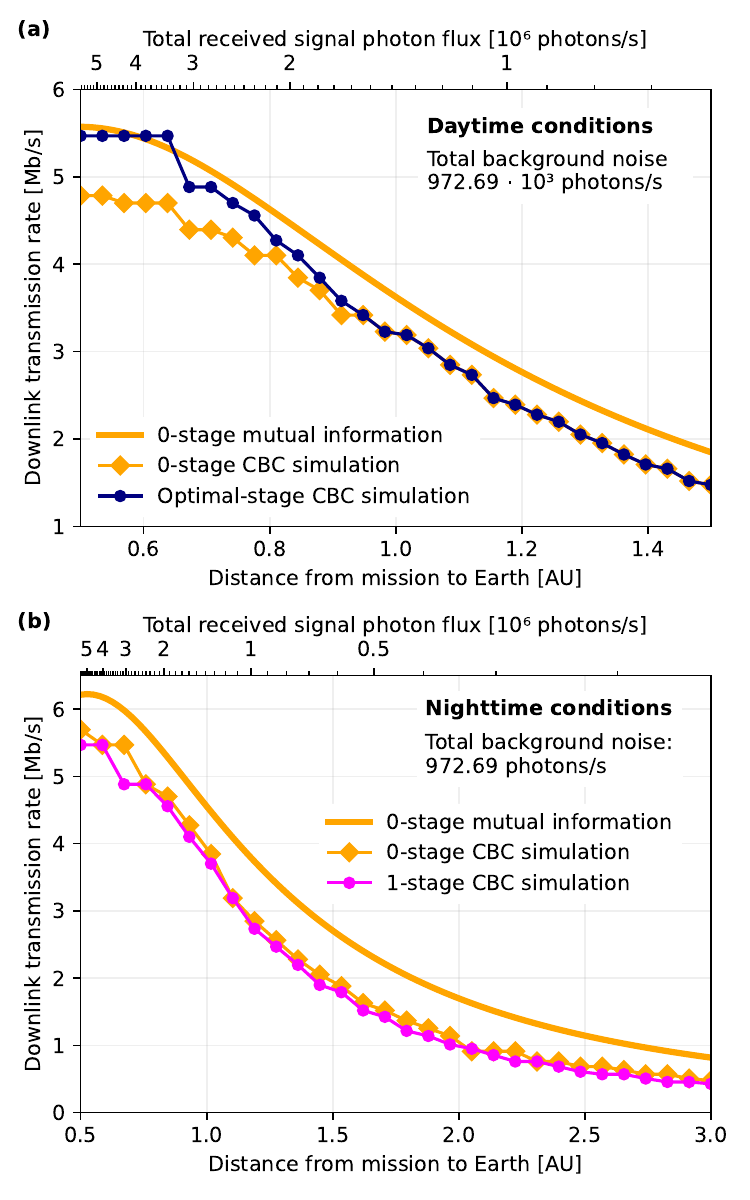}
    \caption{\textbf{SCPPM simulated bitrates for downlink transmission in (a) daytime and (b) nighttime conditions}, depicted as a function of distance from the mission to the Earth-based receiver. Additionally, the total received signal photon fluxes and background noise are depicted. 0-stage indicates a direct detection of each individual beam at the receiving aperture's input. In daytime conditions, running the cascaded CBC scheme for an optimized number of stages is advantageous on distances up to 0.8 AU. In nighttime conditions, already one stage of CBC deteriorates the transmssion rate, indicating that direct detection at the input of each receiving aperture is optimal in conditions of negligible noise.
    }
    \label{fig:realcomm}
\end{figure}

In the simulations we adopted the \emph{Psyche} $\to$ Helmos downlink conditions with parameter values listed in Table~\ref{tab:communication}. Specifically, because of the ambiguity in the distance-unrelated loss factor $\eta_{\text{tot}}$ discussed above, we assumed a pessimistic scenario and included the attenuating factors of ref.~\cite{ivanovReviewDeepSpace2025a} totaling to 14.07 dB plus a 4 dB link margin of ref.~\cite{rielanderESAGroundInfrastructure2023b}. The background noise power spatial density was taken from ref.~\cite{ivanovReviewDeepSpace2025a}. For CBC, we replaced the large 2.28 m receive aperture by 128 small apertures, each of $20$ cm diameter, resulting in a comparable reception area. We run the simulations under both nighttime and daytime background noise values, with weak atmospheric turbulence inducing fluctuation in phase and intensity of the detected beams. 

The obtained SCPPM bitrates are depicted in Fig.~\ref{fig:realcomm} as a function of the distance from the mission to Earth. Alongside we plot the values of mutual information, the upper bound to the attainable coded transmission rate described in Appendix~\ref{app:MI}, which we obtain under the assumption that the beams are directly detected at the input of the cascade and therefore denote as $0$-stage mutual information. The roughly $\sim10\%$ difference between the mutual information and simulated rates comes from finite SCPPM code efficiency. It is seen in Fig.~\ref{fig:realcomm}(a) that when noise becomes a significant factor, as in daytime communication, the stage-optimized cascaded CBC approach provides a clear advantage over the single-aperture configuration up to $0.8$~AU. This is a consequence of the CBC ability to increase the SNR at each stage of combination. When the received signal strength drops to a level comparable with the noise, the CBC efficiency decreases significantly, meaning that for larger distances the transmission rate becomes equal to that offered by simple direct detection of all the beams, comparable to that of a single large telescope. Under nighttime conditions depicted in Fig.~\ref{fig:realcomm}(b) the optimal detection strategy is to simply detect the signal directly in each aperture (at 0-stage), whereas already the 1-stage coherent combination lowers the attainable transmission rate due to the noise being negligible and the primary effect of each subsequent stage being a deterioration of the total signal intensity.

\section{Conclusions}
In this work, we demonstrated that the concept of cascaded coherent beam combination can potentially provide a new approach to receiving photon-starved optical signals. We investigated the efficiency of cascaded CBC in terms of two main parameters: phase diffusion and background noise strength. We established two practical rules that must be satisfied to efficiently combine multiple beams coherently. First of all, in order to ensure a satisfactory total combining efficiency the intensity collected by a single CBC aperture must be sufficiently high relative to the diffusion strength so that the efficiency of combining two beams reaches at least $90\%$. Secondly, the background noise must be weaker than the signal received by a single aperture. If either of these conditions is not met, the cascaded CBC overall efficiency decreases significantly at each combination stage.

In Sec.~\ref{sec:diameters}, we discussed the implications of choosing different aperture diameters and their number. We showed that there exists a lower limit on the aperture diameter, that comes from the first rule mentioned above. We explained that if using telescopes with smaller diameters leads to even a few percent reduction in the efficiency of combining two beams, it can have a substantial impact on the overall efficiency, especially since smaller telescopes require more combination stages. On the other hand, we noted that it would be highly beneficial if the telescope diameter was comparable or smaller than the Fried parameter, which can reach over $20$~cm.  Furthermore, we showed that, for realistic parameters for communication with \emph{Psyche} mission, an array of $128$ telescopes with $20$~cm diameters represents a well-balanced configuration, giving comparable results to the $2.28$~m telescope from Helmos Observatory. Importantly, the presented approach can be also in principle combined with adaptive optics, which would offer larger individual apertures in a CBC array, increasing the signal strength. This may be advantegous in the case of even weaker signals typical for very large distances. Note also that the cost of adaptive optics scales very fast with the size of the aperture \cite{belleScalingRelationshipTelescope2004}, meaning potential economical benefits from replacing large apertures by a set of small ones also equipped with AO.

A major advantage of using CBC in real time is that it produces a coherent output beam, which enables the application of advanced post-processing techniques. This feature may be even more valuable for applications other than PPM-based communication. In our case, however, without affecting the final output beam, one can extract additional information from photon counts in the dark ports at each CBC stage. We showed that this information in a simple way can be used for message decoding by including photons detected in the dark ports after a threshold stage onward. Moreover, one can consider an optimal strategy for utilizing this information, since with each CBC stage the noise level in the output beam is reduced by half. Consequently, the probability that a photon detected in the dark port originates from the signal increases geometrically with the number of stages. Therefore, in principle, this information can be used for optimized message decoding.

We examined a representative use case by simulating PPM-based communication under realistic conditions, using parameters inspired by reports on the \emph{Psyche} mission. We showed that for nighttime DSOC communication, the CBC approach achieves performance comparable to that of a single large aperture. This is a promising result for the practical implementation, as it may offer greater scalability, easier maintenance, and potential economic benefits. The benefits are even more clearly visible under daytime conditions, where background noise becomes a significant factor limiting the transmission. We showed that, due to the high noise robustness of CBC, a clear advantage of CBC over single-large aperture approach is maintained over a range of transmitter distances from Earth. This advantage vanishes when the signal power approaches the background noise level.

The above results are promising for the design of ground-based architectures for receiving messages from future deep-space missions, where the expected signals will be extremely weak. It is also worth noting that the concept of CBC receivers is more general and can be applied to other communication protocols. In particular it may useful for coherent optical communication \cite{larssonCoherentCombiningLowpower2022b} or satellite-based quantum key distribution \cite{luMiciusQuantumExperiments2022, dlr217519, hrynykCanadianSpaceAgency2025, diamantiPracticalChallengesQuantum2016}, where signal coherence and robustness to noise are crucial.

\section{Acknowledgments}
\noindent This work was supported by the project “Quantum Optical Technologies” (contract no. FENG.02.01-IP.05-0017/23) carried out within the Measure 2.1 International Research Agendas programme of the Foundation for Polish Science co-financed by the European Union under the European Funds for Smart Economy 2021-2027 (FENG). This research was funded in part by the National Science Centre, Poland grant SONATA 19 No. 2023/51/D/ST7/02171. This work was supported by the Polish Ministry of Education and Science under the ``Quantum strategies in communication through noisy optical channels'' project no. PN/01/0204/2022 carried out within the ``Pearls of Science'' program.
\appendix[Mutual information for PPM transmission] \label{app:MI}
In Section \ref{sec:res_SNR}, mutual information (MI) is used to estimate the upper bound on the information rate in optical communication employing PPM with cascaded CBC as the receiver. Here, we provide the technical details of how this quantity is computed.

Our model is as follows: the transmitter sends a message encoded using PPM symbols of orer $M$. The message is carried by an optical beam that propagates through the atmosphere and is collected by an array of apertures. The beams from all apertures are then coherently combined in a cascaded configuration, producing a final beam that is measured by a binary photodetector. As described in Sec.~\ref{sec:early}, one may also use photocounts from the dark ports at each combination stage. To compute the MI in such case, we fix a threshold stage and include all photocounts from detectors placed in succeeding stages as well as from the final beam. For each PPM time slot, if a photon was detected, we assume it was occupied by a light pulse; otherwise we assume the slot was empty.

For the direct-detection receiver based on Geiger mode photon counting the formula for MI in the PPM based communication with the background noise is given in~\cite{zwolinskiRangeDependenceOptical2018}
\begin{align}
I &= \frac{1}{M} \sum_{k=1}^{M} \Bigg[
\binom{M-1}{k-1} p_c(k)\, \log_2 p_c(k)+ \\
& \binom{M-1}{k} p_e(k)\, \log_2 p_e(k)
- \binom{M}{k} p(k)\, \log_2 p(k) \Bigg],\nonumber \\
p(k) &:= \frac{k}{M} \, p_c(k) + \left( 1 - \frac{k}{M} \right) p_e(k),\\
p_c(k) &:= p_c\, p_b^{\,k-1} (1 - p_b)^{\,M-k},\\
p_e(k) &:= (1 - p_c)\, p_b^{\,k} (1 - p_b)^{\,M-k-1},
\end{align}
where $M$ is the PPM rank, $p_c$ denotes probability of detecting a photon in a PPM time slot occupied by a light pulse, and $p_b$ is probability of detection without in an empty slot. These probabilities are taken from the Poisson distribution and they depend on the intensities of light in the occupied and unoccupied slots, $I_c$ and $I_b$ respectively
\begin{align}
    p_c^\textrm{temp}(I_c) &= 1-\exp{\big(I_c\Delta t_\text{PPM}\big)},\\
    p_b^\textrm{temp}(I_b) &= 1-\exp{\big(I_b\Delta t_\text{PPM}\big)}.
\end{align}
Due to the atmospheric turbulence and the procedure of CBC both $I_c$ and $I_b$ may randomly fluctuate and suffer from outages. Therefore, for fixed conditions we perform a long simulation of cascaded CBC and obtain a numerical distributions of these intensities, $p(I_c)$ and $p(I_b)$ respectively, for every stage of combination. Having that, we assume both transmitter and receiver do not know the current intensities, therefore we average the temporal probabilities $p_c^\textrm{temp}, p_b^\textrm{temp}$ over these distributions
\begin{align}
    p_c &= \int  p_c^\textrm{temp}(I_c) p(I_c)dI_c,\\
    p_b &= \int  p_b^\textrm{temp}(I_b) p(I_b)dI_b,
\end{align}
which allows to compute MI.
\bibliographystyle{IEEEtran}
\bibliography{ppm}
\end{document}